# Large-Scale Experiment on the Importance of Social Learning and Unimodality in the Wisdom of the Crowd

DHAVAL ADJODAH*, SHI KAI CHONG, YAN LENG, PETER KRAFFT and ALEX PENTLAND,
MIT Media Lab, *Foundation Bruno Kessler, Italy

## 1. INTRODUCTION

Using 17420 predictions that we collected from 2037 students, we build on previous research to understand the conditions within which the Wisdom of the Crowd (WoC) improves or worsens as a result of showing individuals the predictions of their peers. We use social learning weights to filter individuals who are better at learning from the crowd to improve accuracy. We also find the novel effect of the importance of the unimodality of the social information shown to individuals.

## 2. LITERATURE REVIEW

Although harnessing the WoC has had some fascinating sucesses such as in predicting the reproducibility of scientific research [Dreber et al. 2015] and has been shown to promote cooperative behavior [Szolnoki et al. 2012], there is still significant disagreement in the literature as to the role of social learning and influence. Our paper aims to investigate the two opposite claims that social influence can improve or worsen the WoC by investigating the parameter spaces within which these opposite results arise. We build on the results of many: [Lorenz et al. 2011] argue that social influence (when individuals are exposed to the crowd's estimates) decreases the diversity of opinions enough that it can cause significant deviations from the pre-influenced crowd prediction, and [Muchnik et al. 2013] finds a similar result in the domain of online ratings. On the other hand, [Madirolas and de Polavieja 2015] argues that although [Lorenz et al. 2011] might be correct in terms of the accuracy of the whole crowd, it is still possible to recover individuals who are resistant to influence by estimating social weights that indicate their susceptibility to influence. Surprisingly, an aggregate over these resistant individuals can then lead to improvements even beyond the original pre-influence crowd prediction. [Kerckhove et al. 2016] use a similar approach by estimating a parameter of influenceability to predict how much individuals revise their judgements after seeing the opinions of others, while [Soll and Larrick 2009] call it a 'weight on self'.

## 3. DATA

We collected a dataset of unprecedented size and depth that we detail here. Over the span of 6 months, we ran 6 sequential individual rounds and got 2037 students (from several online classes) to make 17420 predictions of the price of a real financial instrument (e.g. the S&P 500) over a period open for 3 weeks for each round - students can come back any time to make predictions again. We collaborated with the crowdsourced stock rating website Vetr.com and designed the prediction process as such: students go to our page, are given some minimal information on the instrument (such as the current price trajectory), and are asked to predict the closing price of a specific instrument on a chosen date along with an obligatory confidence rating (no default value) and their (optional) personal reasoning for the predicted price - this is what we call the pre-social prediction. As soon as this pre-social prediction is submitted, a pop-up is displayed where the students are shown a detailed histogram of the crowd's predictions and asked if they would like to change their prediction: they can put a new number, adopt the current mean of the social histogram, or decide to not update their predictions - all three options are what we call the post-social predictions.

## 4. ANALYSIS & RESULTS

### 4.1 Social Weight

Under the model proposed by [Madirolas and de Polavieja 2015], for each pair of predictions made by a user in a round, the pre-social prediction $P_{pre}$ and post-social prediction $P_{post}$ are related via the following equation $log(P_{post}) = (1 - SW)log(P_{pre}) + (SW)log(\bar{p})$ where $SW$ is the social weight, a proportionality constant associated with the level of social influence on a person's decision and $\bar{p}$ is the geometric mean of the predictions of others that was shown to the user. From the previous equation, we can calculate SW, $SW = \frac{log(P_{post}) - log(P_{pre})}{log(\bar{p}) - log(P_{pre})}$

As shown in Figure 1A, the biggest predictor of improvement is the sign of SW: positive social weights means that the individuals chose a value between their pre-social prediction and the mean of the crowd, negative SW means they go in the opposite direction from the crowd. When judged against the true value, moving towards the WoC is seen to improve the WoC.

We then select a subset $S_\alpha$ of the predictions collected such that all predictions lie within $i \in S$, $0 \leq SW_i \leq \alpha$ if $\alpha > 0$ and $\alpha \leq SW_i \leq 0$ if $\alpha < 0$, i.e we filter predictions by their social weights from 0 to $\alpha$. For each round $i$, we first calculate the arithmetic mean of all the pre-social predictions $P_{i,pre}(\alpha)$ and post-social predictions $P_{i,post}(\alpha)$ in the subset defined by $\alpha$. We then calculate the error from the truth of $P_{i,pre}(\alpha)$ and $P_{i,post}(\alpha)$ each, and define the improvement $I_i(\alpha) = E_{i,pre} - E_{i,post}$. The mean improvement across all rounds is $I(\alpha) = \frac{\sum_i I_i(\alpha)}{\sum_i (1)}$. Finally, the distribution of $I(\alpha)$ is obtained by bootstrapping with replacement on the subset $S_\alpha$ 100 times for each value of $\alpha$. $I(\alpha)$ can then be used to investigate the relative effect of the WoC on subsets based on their influencability by varying $\alpha$ between -1 to 1. As shown in Figure 1B, more positive SW improves the WoC and more negative SW worsen the WoC. This would help explain why some studies have seen the opposite effect of social influence on the WoC: it depends on the SW of the people being sampled.

### 4.2 Unimodality

The distribution of predictions that was shown to the users prior to their post prediction was investigated for unimodality to test if seeing a clear peak in the histogram of the predictions of the crowd helps people predict better. A Hartigans' dip test [Hartigan and Hartigan 1985] was used to test if the histogram that was shown to the user was unimodal. The post-social prediction was flagged as non-unimodal if the p-value of the test on the histogram that was shown was less than 0.05, and unimodal if otherwise.

For each round $i$, we grouped the pairs of pre-social and post-social predictions by each user $j$ and divided these predictions into two mutually exclusive subsets $k \in \{uni, non\_uni\}$ based on the unimodality of the social information given to the user at the time of prediction. For each subset $S_{i,j,k}$, we calculated their average improvement, $I_{i,j,k}$, then investigated the proportion of users whose performance improved ( $I_{i,j,k} > 0$), worsened ($I_{i,j,k} < 0$), or remained the same ($I_{i,j,k} = 0$). The bar chart in Figure 2C shows the result across all rounds and users for both unimodal and nonunimodal social information. A p-value greater than 0.05 of a 1-sample proportions test without continuity correction implies that there is no significant improvement, as in the case of the non-unimodal social information - i.e. the proportion of improved and worsened predictions are not statistically different. This suggests that people predict worse when they cannot find a clear signal of what the crowd's belief is. The plot in Figure 2D shows the individual prediction-wise improvements $I_{i,j,k}$ in each of the subsets, sorted in a descending order. Also shown is how the the average improvement of predictions under unimodality is greater than under non-unimodality.

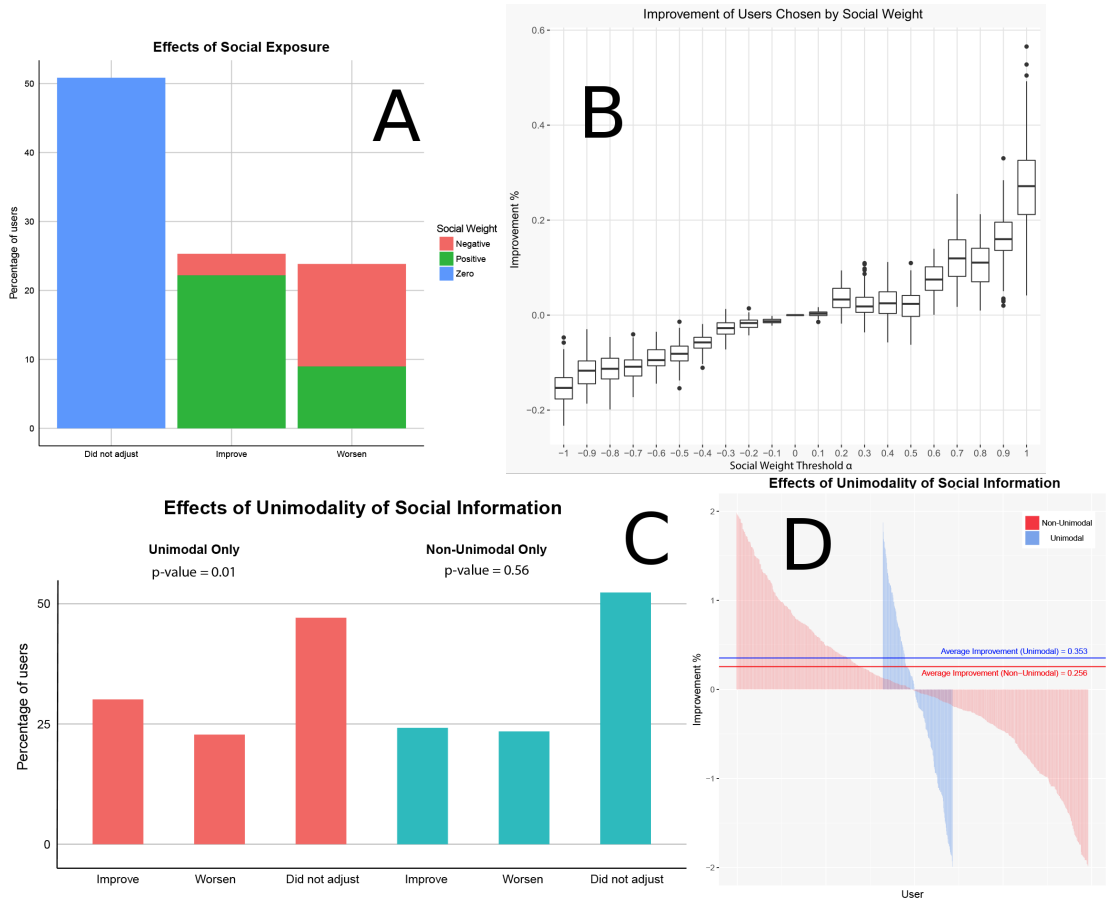

Fig. 1. **A**: For the predictions that are changed after a student sees the histogram of their peers, most of the ones that are worse are associated with a negative social weight, and vice-versa. **B**: The WoC is improved as predictions with increasingly positive SW (the threshold $\alpha$) are selected, and vice-versa. **C**: Users who see a unimodal histogram of their peers statistically improve their predictions; for non-unimodal histograms, the proportion of improved and worsened predictions are not statistically different. **D**: Unimodal predictions have a greater average positive improvement than non-unimodal ones; sorted in a descending order.

## 5. CONCLUSION

First, we investigated the role of social influence by calculating social weights and using them to filter for individuals with higher or lower social weights. As we show, the WoC can be improved through social exposure (when showing individuals the histogram of predictions of their peers) by carefully selecting individuals of positive social weight. Secondly, although showing individuals the histogram of their peers can be beneficial, it strongly depends on whether there is a clear peak (unimodality) in the histogram they are seeing. To our knowledge, this is a novel contribution.